\documentclass[aps,prb,reprint,superscriptaddress,showpacs,amsmath,amssymb]{revtex4-1}

\usepackage{color, bm}
\usepackage[final]{graphicx}
\usepackage{subfigure}
\usepackage{nicefrac}
\usepackage{siunitx}
\usepackage[version=3]{mhchem}

\newcommand{\eq}[1]{\begin{align} #1 \end{align}}
\newcommand{\dif}{\, \mathrm d}
\renewcommand{\vec}{\mathbf}
\newcommand{\ket}[1]{\ensuremath{\left| #1 \right>}}
\newcommand{\braket}[2]{\left< #1 | #2 \right>}
\newcommand{\average}[1]{\left\langle #1 \right\rangle}

\newcommand{\z}{\hat Z}
\newcommand{\ve}{\vec E}
\newcommand{\vh}{\vec H}
\newcommand{\nm}[1]{\SI{#1}{\nano\metre}}
\DeclareMathOperator{\real}{Re}

\begin{document}
\title{Impedance generalization for plasmonic waveguides beyond the lumped circuit model}
\author{Thomas Kaiser}
\email[Corresponding author: ]{Thomas.Kaiser.1@uni-jena.de}

\affiliation{Institute of Applied Physics, Abbe Center of Photonics, Friedrich-Schiller-Universit\"at Jena, Max-Wien-Platz 1, 07743 Jena, Germany}
\author{Shakeeb Bin Hasan}
\affiliation{Institute of Condensed Matter Theory and Solid State Optics, Abbe Center of Photonics, Friedrich-Schiller-Universit\"at Jena, Max-Wien-Platz 1, 07743 Jena, Germany}
\author{Thomas Paul}
\affiliation{Fraunhofer Institute of Applied Optics and Precision Engineering, Albert-Einstein-Stra{\ss}e 7, 07745 Jena, Germany}
\author{Thomas Pertsch}
\affiliation{Institute of Applied Physics, Abbe Center of Photonics, Friedrich-Schiller-Universit\"at Jena, Max-Wien-Platz 1, 07743 Jena, Germany}
\author{Carsten Rockstuhl}
\affiliation{Institute of Condensed Matter Theory and Solid State Optics, Abbe Center of Photonics, Friedrich-Schiller-Universit\"at Jena, Max-Wien-Platz 1, 07743 Jena, Germany}
\date{\today}


\begin{abstract}
We analytically derive a rigorous expression for the relative impedance ratio between two photonic structures based on their electromagnetic interaction.
Our approach generalizes the physical meaning of the impedance to a measure for the reciprocity-based overlap of eigenmodes.
The consistence with known cases in the radiofrequency and optical domain is shown.
The analysis reveals where the applicability of simple circuit parameters ends and how the impedance can be interpreted beyond this point.
We illustrate our approach by successfully describing a Bragg reflector that terminates an insulator-metal-insulator plasmonic waveguide in the near-infrared by our impedance concept.
\end{abstract}

\maketitle

\section{Introduction}
The complexity of electromagnetic phenomena in photonic structures demands for the introduction of simple parameters which allow an easy modeling and engineering of devices.
Among these simple parameters, the impedance is one of the most prominent quantities.
The striking advantage of the impedance concept is that essential properties such as reflection and transmission at the interface between two different structures can be described by simple and well-known relations.
Originating from the radio-frequency domain, many attempts have been made to make use of this concept also in optics and plasmonics by employing analogies to the low-frequency case.
However, a conclusive and rigorous route how the concept could be generalized in cases where no obvious analogy is present has not yet been found.

The impedance concept has a long-standing tradition in various branches of physics. Originally introduced in the late $19^{\rm th}$ century as the ratio of complex voltage and current by Heaviside\cite{Heaviside1888}, the evolution of electromagnetic theory was accompanied by subsequent generalizations of the concept for electromagnetic waves.
As early as 1938, Schelkunoff pointed out that the impedance must be seen as a property of the wave in a medium rather than of the medium alone\cite{Schelkunoff1938}.
Up to the microwave frequency domain, the slow variation of the involved electromagnetic fields allows to simplify the functionality of a device with simple "lumped circuit" quantities such as resistance, capacitance or inductance which constitute the impedance\cite{Collin1992}.
This allowed for unprecedented possibilities to design and understand the physical behavior of devices in many applications for which the impedance is the property of utmost importance.
It is thus not astonishing that several attempts have been made to make use of the concept in optics and plasmonics in a similar way.

The initial meaning of the impedance in electromagnetic wave theory was to describe the ratio of the electric and magnetic field strength.
However, this led to the question how the impedance can be generalized to describe more complex photonic structures in which this ratio is not spatially constant.
Especially for waveguiding devices, a confusing situation with mutually contradictory definitions was reached relatively fast\cite{Walker1966}.
The reason for this was that the ratio between the electric and magnetic field is not necessarily constant over the cross-section in most devices.
This  caused several suggestions for definitions using averaged or integrated fields which were heuristically proposed for a manifold of photonic structures, not only waveguides. 
Examples include the area of photonic crystals where several such heuristic approaches existed\cite{Boscolo2002, Biswas2004, Momeni2007} until it was proven that the so-called Bloch impedance -- the ratio of the surface averaged fields -- was the analytically correct solution, provided that the photonic crystal operates in its fundamental mode\cite{Lu2007, Smigaj2008}.
The same argumentation was shown to be valid for the delicate task of assigning meaningful effective medium parameters to metamaterials\cite{Simovski2011, Paul2011}.
If the excitation of higher order modes becomes important, Lawrence et al. showed that a higher-dimensional impedance matrix becomes necessary\cite{Lawrence2008, Lawrence2009}.

Especially in plasmonics, the design of integrated circuits that relies on simple parameters such as the impedance is highly desired.
Substantial work has been done by Engheta, Al\`u and coworkers on the use of traditional radio frequency concepts for nanostructures\cite{Engheta2005, Engheta2007, Alu2008, Alu2008a}.
These concepts employ the analogy between conducting and displacement current to establish an impedance understood in analogy to the traditional definition as ratio between "optical voltage" and "optical current".
Such an attempt is possible as long as the electromagnetic response closely resembles the solution in the quasi-static limit\cite{Biagioni2012}.
The similarity between a plasmonic metal-insulator-metal waveguide and a waveguide at microwave frequencies, for instance, facilitated the use of traditional impedance definitions for this kind of structures in plasmonics\cite{Veronis2005a, Cai2010, Nejati2012}.
However, this approach generally requires that lumped circuit parameters can be found by analogies to the radio-frequency domain.
A conclusive route for a generalization beyond this assumption has not yet been presented although attempts have been made to understand the impedance in a broader sense, e.g. for quantum emitters\cite{Greffet2010}.

It has been pointed out by Hecht and coworkers that an experimentally relevant impedance definition must correctly describe the reflection that occurs at the boundary between \emph{two} different structures\cite{Huang2009, Biagioni2012}.
From the practical point of view, this means that the impedance of a certain structure is described relative to a reference structure from which it is illuminated or excited.

\begin{figure}
	\centering
		\subfigure[~sketch of the test geometry]{\includegraphics{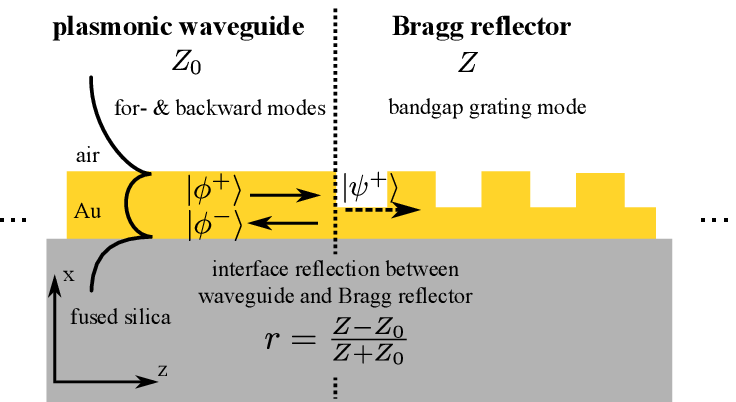}} \\
		\subfigure[~equivalent circuit model]{\includegraphics{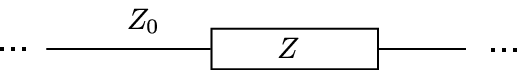}} \\
	\caption{(Color online) (a) Sketch of the test geometry considered in this paper. The fundamental mode of a plasmonic insulator-metal-insulator waveguide is reflected by a Bragg reflector. (b) The situation is characterized in terms of the relative impedance between the two structures as in an equivalent circuit model.}
	\label{fig:sketch_plasmonic_waveguide}
\end{figure}

In this contribution, we bridge the gap between the fundamental and the practical approach by rigorously analyzing the reflection at an interface of impedance discontinuity, i.e. between two different structures.
The basic structure we consider is shown in Fig.~\ref{fig:sketch_plasmonic_waveguide}(a).
A plasmonic waveguide, characterized by a referential impedance $Z_0$, illuminates another photonic structure. 
In our test example, this will be a Bragg reflector, i.e. periodic corrugations in a metal film, which we intend to describe by an impedance $Z$.
In the sense of circuit theory, this problem can be considered as part of a photonic network as shown in Fig.~\ref{fig:sketch_plasmonic_waveguide}(b).
Since our approach is based on a decomposition of the electromagnetic fields into eigenmodes, the results and conclusions will not be limited to this particular case.
Rather, we chose this example because it displays the main difficulties of the task: the relatively strong loss in the metal that prevents the use of many radio-frequency derivations and the open boundary condition that makes it impossible to find suitable analogies to voltage and current.

The paper is organized as follows.
In Section~\ref{sec_methods}, the unconjugated reciprocity framework is set up to solve for the interface reflection coefficients of the different electromagnetic eigenmodes.
This rigorous expression is linked to the relative impedance of the discontinuity between the two structures in Sec.~\ref{sec_results}.A which yields a very general expression for the modes' impedances.
The observed entanglement between them is discussed and the formula is specialized to the practical important case of a single mode interaction.
In Sec.~\ref{sec_results}.B, we show that our analysis is consistent with previous results and reveals that the modal symmetries of the structure are the key point that allows absolute impedance definitions in these cases.
We apply our findings to the case of the plasmonic insulator-metal-insulator waveguide with a Bragg mirror termination as a test example in the last part of Sec.~\ref{sec_results}.
We conclude our work by discussing the implications of our findings.

\section{Theoretical analysis}\label{sec_methods}

The purpose of introducing the impedance concept to various branches of physics  has without doubt been the desire to describe similar processes in the same conceptual way.
The quantities that are ultimately associated to the impedance in that sense are the transmission and reflection at an interface of impedance discontinuity.
They are given by the well-known expressions
\eq{\label{eq_rtz}
	r = \frac{Z - Z_0}{Z + Z_0} , \qquad
	t = \frac{2Z}{Z + Z_0} ,
}
if one assigns the impedances $Z$ and $Z_0$ to the two structures under consideration \cite{Schelkunoff1938}.

For a meaningful impedance definition, eqs. \eqref{eq_rtz} should hold. 
Our attempt for the derivation of a generalized impedance for plasmonics will be an inversion of the above equations, i.e. an retrieval of the impedance from reflection and transmission coefficients.
An important fact is that the above equations depend solely on the relative impedance ratio $\z = Z/Z_0$ of the two structures and that they obey $r + 1 = t$.
This means that the relative impedance can already be derived from either the reflection or the transmission while the other quantity would not give new information.
The individual impedances themselves need to be normalized afterwards in order to be unambiguous.
If we invert the reflection, we find
\eq{\label{eq_zr}
\z = \frac{Z}{Z_0} = \frac{1+r}{1-r} .
}

In order to rigorously derive an impedance for plasmonic waveguides, an analytic expression for the reflection needs to be known.
The framework that provides this knowledge is a decomposition of the electromagnetic fields under consideration into the eigenmodes of the structures that support them.
We denote the $n^\text{th}$ eigenstate of the structure which should be described by the impedance $Z$ by an abstract ket vector \ket{\psi_n}.
This state is described by a concatenation of its tangential electromagnetic field components, i.e. the transverse components of $\ve_n$ and $\vh_n$.
The longitudinal components are then determined by Maxwell's equations.
In the same manner, we describe the modes of the structure from which we illuminate the functional element with \ket{\phi_n}.
In our test scenario in Fig.~\ref{fig:sketch_plasmonic_waveguide}(a), $\{\ket{\phi_n}\}$ are the different bound and radiating modes of the plasmonic waveguide, whereas  $\{\ket{\psi_n}\}$ are the Bloch modes of the periodically corrugated metal film.
We distinguish forward- and backward propagating modes by a superscript plus or minus.
\ket{\phi^+_1}, for instance, represents the fundamental illuminating mode of the plasmonic waveguide.

The virtue of using a modal framework is the mathematical orthogonality of the modes, allowing to unveil basic physical processes and to derive analytical formulas.
It must not be forgotten that the term "orthogonal" refers to a specific inner product.
In electromagnetic theory, two forms of inner products are derived from the reciprocity relation\cite{Snyder1983}.
The first version is based on conjugated reciprocity, involving products of the $\ve \times \vh^*$ form. 
This one is widely used in waveguide theory since the obtained expressions are directly related to the energy flux given by the time-averaged Poynting vector $\average{\vec S} = \nicefrac{1}{2}\,\real\,[\ve\times\vh^*]$. 
However, this version has a substantial drawback. It is only valid for loss-less structures. 
This prevents the use in plasmonics where the loss of metals at optical frequencies must be taken into account.
Hence, an appropriate inner product for modal analysis in plasmonics is only given by the unconjugated version of the reciprocity theorem. 
This reads as\cite{Lalanne2005b,Snyder1983}
\eq{\label{eq_inner_product}
\braket{\phi_m}{\phi_n} = \iint_{\mathbb{R}^2} 
	\bigl[\ve_n\times\vh_m - \ve_m\times\vh_n \bigr]\dif\vec A ,
}
where $\vec A$ points into the forward propagation direction which we choose to be the $z$-axis.
The definition is applicable to modes of $z$-invariant waveguides, as well as to the Bloch modes of periodic waveguides\cite{Lecamp2007a}.
The orthogonality relations are explicitly given as
\eq{\label{eq_orthogonality}
\begin{array}{rclrcl}
	\braket{\phi^+_m}{\phi^+_n} &=& 0 \qquad\qquad & \braket{\phi^-_m}{\phi^+_n} &=& C_{m}\delta_{mn} \\
	\braket{\phi^-_m}{\phi^-_n} &=& 0 & \braket{\phi^+_m}{\phi^-_n} &=& -C_{m}\delta_{mn}
\end{array}
 ,
}
with a normalization constant $C_m$.
For a detailed derivation and discussion of the orthogonality relations we refer to the literature \cite{Lecamp2007a, Smigaj2011, Paul2011}. 

With these relations at hand, the reflection at the interface can be decomposed into eigenmodes of both structures.
Let $r_{mn}$ be the reflection into the $m^{\rm th}$ backward mode caused by the $n^{\rm th}$ forward mode and similar for $t_{mn}$.
Maxwell's equations demand the continuity of the transverse electromagnetic fields across the interface which reads as
\eq{\label{eq_modal_decomposition}
\ket{\phi^+_n} + \sum_m r_{mn}  \ket{\phi^-_m} 
	= \sum_m  t_{mn} \ket{\Psi^+_m} .
}
An arbitrary illuminating field \ket{\phi^+} can be decomposed into forward modes as well
\eq{\label{eq_illumination}
\ket{\phi^+} = \sum_n c_n \ket{\phi^+_n} ,
}
where $c_n$ are the excitation coefficients.
By successively projecting \eqref{eq_modal_decomposition} onto modes $\{\ket{\psi^+_m}\}$ and introducing a matrix formulation, it is possible to solve for the reflection matrix
\eq{\label{eq_rin}
 \hat r = - \hat P^{-1} \,\hat Q,
}
where $\hat P$ and $\hat Q$ are matrices with the elements
\eq{\label{eq_pq}
P_{mn} = \braket{\psi_m^+}{\phi_n^-}, \qquad
Q_{mn} = \braket{\psi_m^+}{\phi_n^+}.
}
With these very general expressions at hand, one can analytically introduce the impedance.

\section{Results}\label{sec_results}
\subsection{Relative Impedance formula}

From the above derivation it becomes immediately clear that a matrix impedance will inevitably become necessary to describe the coupling of all modes analytically exact.
The elements of such an entity can be found by using the elementary intermodal reflections given in \eqref{eq_rin} together with \eqref{eq_zr} to obtain
\eq{\label{eq_zmn}
\hat Z_{mn} = \frac{1 - \sum\limits_k (\hat P^{-1})_{mk}Q_{kn}}
									 {1 + \sum\limits_k (\hat P^{-1})_{mk}Q_{kn}} .
}
This expression describes the general form of a relative impedance between two eigenmodes $m$ and $n$ of two different structures and reflects the first main result of this paper.
It unveils also the main problem that has hindered a successful introduction of a generalized impedance in the past.
First, the scalar character of the impedance is lost as soon as the interaction becomes multimode.
Second, from the sum in \eqref{eq_zmn} and the definitions of the matrices $\hat P$ and $\hat Q$ in \eqref{eq_pq}, it becomes obvious that the entire modal sets of the two structures will contribute to a single value $\hat Z_{mn}$.
Yet, information about the full mode spectra of both structures (or at least a numerically feasible subset) are necessary since all modal contributions are entangled.
The total reflection into the $m^{\rm th}$ backward mode 
\eq{r_m = \sum_n c_n r_{mn}}
is obtained by taking the coefficients $c_n$ of the illumination mode spectrum \eqref{eq_illumination} into account. This reads as
\eq{\label{eq_rnz}
r_m = \sum_n c_n \frac{\hat Z_{mn} - 1}{\hat Z_{mn} + 1} .
}

When only few modes play a role, it is  possible to arrive at comparatively compact matrices.
This path was followed in works on photonic crystal anti-reflection coatings of specific lattice geometries, using plane-waves as the eigenmode basis\cite{Lawrence2008, Lawrence2009}.
Further analytic expressions might be derived from \eqref{eq_zmn} for devices where the involved mode sets are manageable.


However, there cannot be any doubt that the strength of the impedance concept, namely greatly simplifying the engineering and physical understanding of structures, is only present when the two structures are (at least approximately) monomode.
Only in this case, a meaningful scalar impedance can be introduced.
Fortunately, this situation is the preferred scenario in many applications in integrated optics and plasmonics.

In the above derivation, we can then omit the mode indices. $\hat P^{-1}$ will become a scalar quantity.
Using the definition of the inner product \eqref{eq_inner_product}, the relative impedance of the two structures then reads as
\eq{\label{eq_z}
\frac{Z}{Z_0} = 
	\frac{\displaystyle \iint_{\mathbb R^2}
		(\ve_0^- - \ve^+_0)\times\vh^+ - \ve^+\times(\vh_0^- - \vh^+_0) \dif\vec A}
			{\displaystyle\iint_{\mathbb R^2} 
		(\ve_0^- + \ve^+_0)\times\vh^+ - \ve^+\times(\vh_0^- + \vh^+_0) \dif\vec A} ,
}
where we have denoted the electromagnetic fields in the reference structure by an index $0$. 
Equation \eqref{eq_z} represents the most general case for a scalar impedance ratio based on the involved fundamental modes.
Certain present symmetries of the structures may be exploited to further simplify the expression, as will be shown hereafter.
It is especially fully valid also for lossy structures and thus well applicable to plasmonic devices.

\subsection{Relation to previous concepts}

Before demonstrating its performance in plasmonics, it is necessary to show that \eqref{eq_z} is consistent with already known cases where impedances were derived in a different way.
For an homogeneous space for instance, an eigenmode basis are plane waves where the transverse field components fulfill the symmetry relation\cite{Lecamp2007a} $\left(\ve_0^-\right)_\perp = \left(\ve_0^+\right)_\perp, \left(\vh_0^-\right)_\perp = -\left(\vh_0^+\right)_\perp$.
The expression then indeed reproduces the known classical result $Z = E/H$, where $E$ and $H$ are the magnitudes of the electromagnetic fields.
If we consider a homogeneous medium with permitivity $\epsilon$ and permeability $\mu$ that is illuminated normally from free space, this coincides with $Z = \sqrt{\mu/\epsilon}$ and $Z_0 = \sqrt{\epsilon_0/\mu_0} \approx \SI{377}{\ohm}$ which is commonly understood as the "intrinsic" impedance of the medium and free space (although it depends on the electromagnetic wave propagating in it).

The next more complicated example is a photonic crystal illuminated by a plane wave.
This case is described by the Bloch impedance $Z_B = \average{E_B} / \average{H_B}$, where $\average{E_B}$ and $\average{H_B}$ are the interface-averaged fields of the pseudo-periodic Bloch modes.
By inserting the plane wave as a reference and exploiting again the plane wave symmetry, it becomes possible to pull the magnitudes $E_0$ and $H_0$ out of the integrals in \eqref{eq_z}.
Taking into account the periodic symmetry of the photonic crystal, the remaining integrals over $\ve^+$ and $\vh^+$ can be taken over the periodic unit cell\cite{Paul2011}. 
Indeed, $Z$ shrinks to the ratio of the averaged electric and magnetic fields in this case and becomes identical to the Bloch impedance.

\begin{figure}
	\centering
		\includegraphics{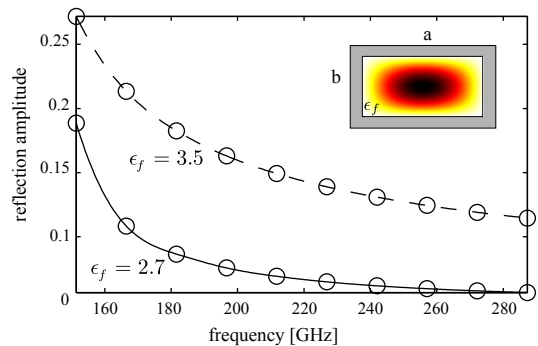}
	\caption{(Color online) Reflection of the fundamental mode of a section with $\epsilon_f = 1$ in a rectangular microwave waveguide ($a=\SI{1}{mm}, b=\SI{0.5}{mm}$) at a section of different dielectric filling $\epsilon_f\neq 1$. Lines represent the rigorous result obtained by COMSOL Multiphysics whereas circles show the results using the impedance framework. The inset shows the geometry and the fundamental mode profile.}
	\label{fig:microwave}
\end{figure}

The treatment of waveguiding devices in general is more sophisticated since none of the involved modes is a pure plane wave.
The inhomogeneous field profiles do not allow to pull contributions out of the integral a priori, but again modal symmetries may be exploited for special cases.

To illustrate this and point out the differences to plasmonics, we consider a rectangular waveguide at microwave frequencies first. 
It consists of a dielectric filling (permitivity $\epsilon_f$) surrounded by a perfect electric conductor of rectangular shape with dimensions $a$ and $b$ as shown in the inset of Fig.~\ref{fig:microwave}.
We investigate the reflection and transmission that occurs when a section with dielectric filling $\epsilon_f\neq1$ is attached to an air-filled section with $\epsilon_f=1$.
Figure~\ref{fig:microwave} shows the results for the reflection amplitude when using \eqref{eq_rnz} together with our impedance definition \eqref{eq_z} in comparison to rigorous results obtained by COMSOL Multiphysics.
The perfect agreement can easily be understood when the analytic properties of the modes are analyzed.

The eigenmodes of this waveguide are separable functions in $x$ and $y$ direction. 
Assuming TE polarization for instance, the forward modes have the explicit form\cite{Collin1992}
\eq{
	H^+_x(x,y) &= i\frac{\beta_{nm}}{\zeta_{nm}^2} n\frac{\pi}{a}
							\sin\left(n\frac{\pi x}{a}\right) \cos\left(m\frac{\pi y}{b}\right)}
\eq{
	H^+_y(x,y) &= i\frac{\beta_{nm}}{\zeta_{nm}^2} m\frac{\pi}{b}
							\cos\left(n\frac{\pi x}{a}\right) \sin\left(m\frac{\pi y}{b}\right) }
\eq{
	E^+_x(x,y) &= \frac{k_0}{\beta_{nm}} Z_{\rm core} \; H^+_y(x,y)}
\eq{
	E^+_y(x,y) &= - \frac{k_0}{\beta_{nm}} Z_{\rm core} \; H^+_x(x,y) , }
with $\zeta_{nm} = \pi[(n/a)^2 - (n/b)^2]^{\nicefrac{1}{2}}$, $Z_{\rm core} = (\mu_0 / \epsilon_0\epsilon_f)^{\nicefrac 1 2}$ and the propagation constant $\beta_{nm} = [\epsilon_f\omega^2/c^2 - \zeta_{nm}^2]^{\nicefrac{1}{2}}$. 
The spatially varying part of the electromagnetic field consists of trigonometric functions that depend solely on the waveguide size, not on the dielectric filling or the propagation constant.
Consequently, although the modes are not constant in space, the inhomogeneous contributions to both integrals in \eqref{eq_z} cancel out.
What remains is just the difference in propagation constants and dielectric filling.
This specific modal property
is the key point that allows to decouple the modal contributions so that it becomes possible to introduce the (absolute) impedance of each waveguide section separately in the well-known\cite{Collin1992} form $Z_{\rm wg} = Z_{\rm core} [1-\omega_c^2 / \omega^2]^{-\nicefrac{1}{2}}$, where the cut-off frequency $\omega_c = \zeta_{nm}c /\sqrt{\epsilon_f}$ was used.
$Z_{\rm wg}$ in this case again gains the physical meaning of the mixed components amplitude ratio $Z_{\rm wg} = |E_x/H_y| = |E_y/H_x|$.
Yet, we were able to show that the impedance known for waveguides in the microwave frequency domain is a special case of our general framework where certain modal properties allow further simplifications of our expression.

\begin{figure}
	\centering
		\includegraphics{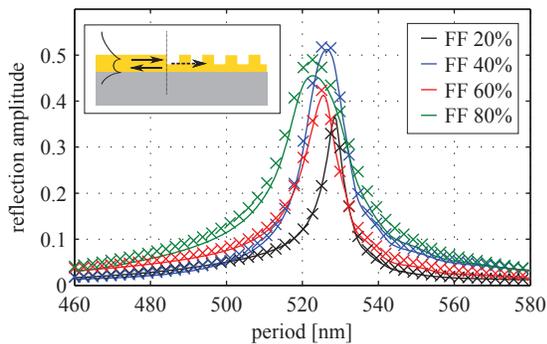}
	\caption{(Color online) Rigorous results (lines) obtained by the aperiodic Fourier-Modal-Method and results of the impedance framework (crosses) for the problem sketched in Fig.~\ref{fig:sketch_plasmonic_waveguide} for different filling factors of the Bragg reflector. The free space wavelength was $\lambda_0 = \nm{1550}$ and the corrugation depth 50\%. A very good agreement is obtained.}
	\label{fig:r_comparison}
\end{figure}

\begin{figure}
	\centering
		\includegraphics{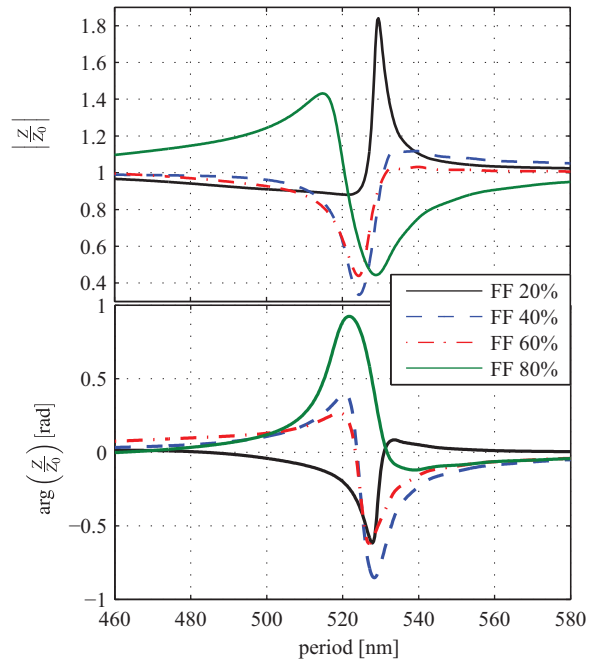}
	\caption{(Color online) Magnitude and phase of the relative impedance for the results of Fig.~\ref{fig:r_comparison}. Depending on the filling factor (FF), the reflection peak has its origin either in a strong mismatch in the former or the latter. }
	\label{fig:impedance}
\end{figure}

\subsection{Application to plasmonics}

In plasmonics, both circumstances of the aforementioned examples are combined, making the task of introducing a meaningful impedance somewhat delicate.
The localized fields hinder the meaningful use of a plane wave basis, similar to the microwave waveguide regime, whereas the high frequency in the optical domain changes the boundary conditions from perfect conducting to an open geometry so that no analogy to the quasi-static limit is present.
Additionally, the presence of possibly strong metal loss within the structure must be taken into account.

We describe the functionality of the plasmonic test device in Fig.~\ref{fig:sketch_plasmonic_waveguide} by our impedance concept in the same way we treated the examples before.
It consists of \nm{30} gold on a fused silica substrate covered by air.
The complex permitivity of gold is fully taken into account\cite{Johnson1972}.
The waveguide is operated in its fundamental mode at telecommunication wavelength $\lambda=\nm{1550}$ and terminated by the Bragg reflector which consists of periodic corrugations in the metal film.
This open geometry is not lumped in the sense of circuit theory and thus not accessible by traditional concepts.

The mode of the plasmonic waveguide is $z$-invariant, possessing exponential decay into the substrate and cladding.
The fundamental mode of the Bragg reflector is of the pseudoperiodic Bloch type and has a photonic band structure.
The nature of the modes does not allow to assume any simplification for \eqref{eq_z} since the problem possesses no exploitable symmetry.
The integral entanglement between the modal fields will thus stay present in the example and cannot be simplified further.
Only a relative impedance description of the reflector with respect to the plasmonic waveguide will be possible whereas the introduction of an absolute impedance for the waveguide or the grating alone is not.
Additionally, the impedance looses its meaning as ratio between the electric and magnetic field but displays a broader interpretation as reciprocity-based overlap between the eigenmodes.

Figure~\ref{fig:r_comparison} shows the results of the analysis.
A prominent reflection peak occurs when the impinging mode is impedance mis-matched to the reflector Bloch mode.
Based on the two modes, we calculated this impedance mismatch according to \eqref{eq_z} and compared the results to rigorous simulations performed by the aperiodic Fourier-Modal-Method \cite{Silberstein2001a, Hugonin2005b, Lecamp2007a}.
Both are found to be in very good agreement which shows the applicability of our concept.
Minor deviations at the resonance position result from a minimal excitation of higher order modes.
Instead of a rigorous simulation, the behavior of the Bragg reflector terminating the plasmonic waveguide can well be predicted just by calculating the impedance mismatch.

Figure~\ref{fig:impedance} shows results for the relative impedance for different filling factors.
Magnitude and phase both contribute to the overall reflection peak.
At the period where the reflection reaches its respective maximum, the physical character of the reflection changes with the filling factor.
For 40\% and 60\% filling factor, a mismatch in magnitude is observed while the phase is matched.
This behavior is reversed for higher or lower filling factors.
This can be understood since the reflector will be more sensitive to the Bloch phase factor if the filling factor is either very low or very high whereas the phase changes less for filling factors around \SI{50}{\%}.
This serves as proof that our impedance description of plasmonic devices can yield additional physical insight.

\begin{figure}
	\centering
		\includegraphics{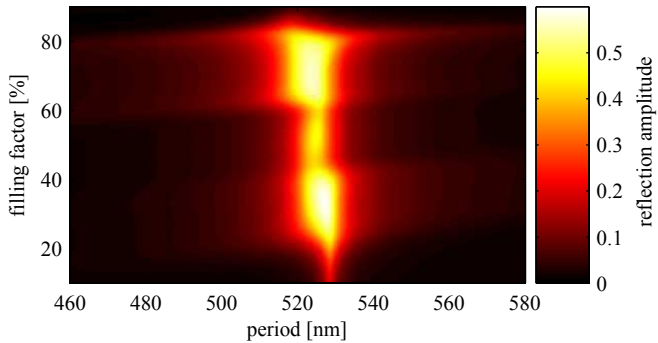}
	\caption{(Color online) Map of the reflection amplitude for different filling factors and periods calculated by the impedance framework. The optimal parameters were found to be 33\% filling factor at a period of \nm{528}.  }
	\label{fig:reflection}
\end{figure}

The filling factor can, in addition to the period, be regarded as optimization parameter for the structure which can be performed by using the impedance mismatch of the reflector with respect to the waveguide.
Instead of running time-consuming rigorous calculations, the impedance framework provides a quick and accurate possibility to scan large datasets as required for this problem.
Figure~\ref{fig:reflection} shows the results of the optimization task.
The best performance is found for a filling factor of 33\% at a period of \nm{528}.

\section{Discussion}\label{sec_discussion}
We derived expressions that generalize the im\-pe\-dance concept for waveguiding devices from the microwave frequency regime to optics and plasmonics.
Our expressions are based on electromagnetic eigenmodes that are excited at the interface of a structure.
Two circumstances have been shown to be the major difficulty for plasmonics.
First, the high operating frequency does not allow for perfect conducting boundary conditions and fields are not lumped in the sense of circuit theory.
Second, the inherent loss of plasmonic structures does not allow to extend the existing concepts for (nearly) loss-less structures.
We overcame both problems by using a rigorous analysis of the modal scattering coefficients at an interface, using the unconjugated version of the reciprocity theorem.
Since a meaningful scalar impedance must correctly reproduce the scattering coefficients, we can invert the found expressions to retrieve the impedance.

Applied to waveguides in the microwave frequency range, we reproduce classically known expression for the impedance.
Moreover, it was shown that in this case, only the specific form of the electromagnetic eigenmodes allows to derive the well-known absolute impedance expression. 

In the optical frequency range, a benchmark of our theory is the extensive work on impedance definitions for photonic crystals where the Bloch impedance was found to consistently describe the plane wave scattering.
Applied to the aforementioned situation, our expression correctly collapses to the Bloch impedance, which is contained as a special case.
The Bloch impedance and our expression therefore appear as different levels of generalization of the impedance concept towards optics.

As a test example of a simple functional element, we considered a plasmonic Bragg reflector, i.e. periodic corrugations in a metal film, which is illuminated by the fundamental mode of an insulator-metal-insulator waveguide.
The results predicted by our formula are in very good agreement with rigorous simulation results.
This proves the potential of our concept as a versatile tool for engineering integrated nanophotonic and plasmonic devices.
Nevertheless our analysis also unveils the limitations of the impedance concept in integrated plasmonics.
We showed that a meaningful scalar impedance can just be introduced if the structure is monomode.
If higher order modes, radiating or leaky modes become important, we also derived a generalization but it comes at the expense of simplicity.
The relative impedance of two modes will in general depend also on all other excited modes in that case.

Our rigorous derivation also reveals that the im\-pedance, especially for plasmonic structures, must in general be regarded as a quantity relative to a reference modal framework.
The introduction of an absolute impedance was shown to require specific modal properties which are not always present in typical plasmonic geometries.
For this situation we were able to derive a very general expression for the relative impedance based on the eigenmodes of the structure and the reference.
This also showed that the impedance changes its meaning from the ratio of the electric and magnetic field to an reciprocity-based overlap of the eigenmodes.

Since the relative impedance matching is usually of much more interest for applications than an absolute impedance value, our derived expression has potential to significantly ease the engineering and physical understanding of plasmonic devices.

\begin{acknowledgments}
The authors gratefully acknowledge financial support by the Federal Ministry of Education and Research (PhoNa), the German Research Foundation 
(MetaFilm, NanoGuide) and the Thuringian Ministry of Education, Science and Culture (MeMa). T.K. thanks P. Lalanne and R. Vogelgesang for useful discussions and some references.
\end{acknowledgments}


%

\end{document}